\documentclass[preprint, aps]{revtex4}

\usepackage{graphicx}

\newcommand{\bq}{\begin{equation}}
\newcommand{\eq}{\end{equation}}
\newcommand{\bn}{\begin{eqnarray}}
\newcommand{\en}{\end{eqnarray}}

\begin{document}

\title{Influence of Magnetic Moment Formation on the Conductance of Coupled Quantum Wires}
\author{V. I. Puller}
\affiliation{Department of Physics, Queens College, the City
University of New York, Flushing, New York 11367, USA}
\author{L. G. Mourokh }
\affiliation{Department of Physics and Engineering Physics,
Stevens Institute of Technology, Hoboken, New Jersey 07030, USA}
\author{J. P. Bird}
\affiliation{Department of Electrical Engineering, University at
Buffalo, the State University of New York, Buffalo, NY 14260-1920,
USA}
\author{Y. Ochiai}
\affiliation{Department of Electronics and Mechanical Engineering,
1-33, Yayoi-cho, Inage-ku, Chiba City, Chiba, 263-8522 Japan}

\begin{abstract}
In this report, we develop a model for the resonant interaction
between a pair of coupled quantum wires, under conditions where
self-consistent effects lead to the formation of a local magnetic
moment in one of the wires. Our analysis is motivated by the
experimental results of Morimoto et al. [Appl. Phys. Lett. 82,
3952 (2003)], who showed that the conductance of one of the
quantum wires exhibits a resonant peak at low temperatures,
whenever the other wire is swept into the regime where
local-moment formation is expected. In order to account for these
observations, we develop a theoretical model for the inter-wire
interaction that calculated the transmission properties of one
(the fixed) wire when the device potential is modified by the
presence of an extra scattering term, arising from the presence of
the local moment in the swept wire. To determine the transmission
coefficients in this system, we derive equations describing the
dynamics of electrons in the swept and fixed wires of the
coupled-wire geometry. Our analysis clearly shows that the
observation of a resonant peak in the conductance of the fixed
wire is correlated to the appearance of additional structure (near
$0.75\cdot$ or $0.25\cdot 2e^2/h$) in the conductance of the swept
wire, in agreement with the experimental results of Morimoto et
al.
\end{abstract}
 \maketitle

\section{Introduction}

The low-temperature conductance of quantum point contacts (QPCs)
is well known to be quantized in units of $2e^2/h$, a phenomenon
that can be explained in terms of a simple transmission (Landauer)
picture in which the influence of electron-electron interactions
is neglected \cite{Datta}. While this model is remarkably
successful in accounting for the observation of conductance steps
at integer units of $2e^2/h$, it is unable to explain the origin
of the additional conductance plateau, observed near $0.7\cdot
2e^2/h$ in numerous experiments. (For an overview of this issue,
see Ref. \cite{Science}.) While many different theoretical models
have been proposed to account for the origins of the 0.7 feature,
there is a wide consensus that it is likely associated with some
novel many-body effect. One of the most convincing explanations
(although not yet commonly accepted) is the development of a net
magnetic moment in the QPC when it is almost pinched off
\cite{Meir1,Berggren,Hirose1}. In our recent work
\cite{Morimoto,Puller,IEEE} we provided experimental and
theoretical support for this idea. The device structure that we
have studied experimentally in Ref. \cite{Morimoto} is shown in
Fig. 1 and was formed in the two-dimensional electron gas of a
GaAs/AlGaAs quantum well. The device  was realized by means of
electron-beam lithography, and lift-off of Ti-Au gates. These
gates were formed on a Hall bar with eight ohmic contacts,
positioned uniformly along its upper and lower edges. In suitable
combinations, these contacts could be used to make four-probe
measurements of the conductance of either wire, or of the quantum
dot itself (as indicated in Fig. 1). Of particular interest here
is the non-local measurement (right panel) that can be made by
measuring the conductance through one (fixed) wire as the gate
voltage ($V_g$) applied to the other (swept) wire is varied. The
key result of our experiment is that as the swept wire pinches
off, a resonant enhancement of the conductance of the fixed wire
is observed. A qualitative theoretical explanation of this
phenomenon was given in Ref. \cite{Puller}. Based on a modified
Anderson Hamiltonian, we showed that the resonant interaction with
the local magnetic moment formed in the swept wire leads to an
additional positive contribution to the density of states of the
fixed wire and, consequently, to an enhancement of its
conductance. While this analysis provides a qualitative
understanding of the resonant interaction between the quantum
wires, the tunnel matrix elements involved in the Anderson
Hamiltonian are generally unknown and have thus far been used as
fitting parameters. In addition, the influence of the specific
device geometry on these matrix elements has thus far been
neglected, even though geometry-related effects are known to be
important for the description of scattering in one-dimensional
structures \cite{Gurvitz1,Stone}. To overcome these shortcomings,
in this paper we present a more comprehensive theory for the
electron dynamics in the coupled-wire system of Fig. 1 and attempt
to calculate the amplitude of the resonant inter-wire interaction
without the assumption of the localized state formed in the swept
wire. To this end, we further develop the approach introduced in
our previous paper \cite{IEEE}, where the conductance of a single
quantum wire was determined. In this work, we obtained such
features as an additional $0.75\cdot 2e^2/h$ plateau for
ferromagnetic coupling between the local moment in the QPC and the
conducting electrons and a $0.25\cdot 2e^2/h$ plateau for the
antiferromagnetic coupling, in agreement to the research of Refs.
\cite{Rejec,Flambaum}. In the present paper we calculate the
single-electron transmission properties of the fixed wire in a
device potential that is modified by the presence of an extra
scattering term, arising from the presence of a local magnetic
moment in the swept wire. The formulation of this idea is given in
Section II, where we derive equations describing the dynamics of
electrons in the swept and fixed wires. In Section III, we
determine the transmission coefficient and conductance for the
fixed wire and compare these expressions to the results of Ref.
\cite{IEEE}. In particular, we show that an additional peak in the
conductance of the fixed wire is correlated to the appearance of
additional plateaus (at $0.75\cdot 2e^2/h$ or $0.25\cdot 2e^2/h$)
in the conductance of the swept wire in agreement with the
experimental results of Ref. \cite{Morimoto}. The conclusions are
presented in Section IV.

\section{Electron modes in the coupled quantum wire structure}

We start our description of electron dynamics in the coupled
quantum wire structure from the following single-particle
Hamiltonian \bq
\hat{H}_{0}=K_{x}+K_{y}+U(x)+W(y)+V(x,y)-J(x,y)\hat{\vec{\sigma}}\cdot\hat{\vec{S}},
\label{eq:F1} \eq where $K_{x}$ and $K_{y}$ are the kinetic energy
operators for an electron localized in the 2D plane, $W(y)$ is the
double-well potential describing the two quantum wires (Fig. 2,
center panel), $V(x,y)$ is the potential of the tunnelling channel
connecting the two wires (Fig. 2, right panel), and $U(x)$
describes the smooth bottleneck shape of the quantum wire
channels. The last term simulates exchange coupling between the
conductance electrons (Pauli matrices $\hat{\vec{\sigma}}$) and
the local moment, $\hat{\vec{S}}$, which is assumed to be a
spin-1/2 magnetic moment with $J(x,y) $ as the
coordinate-dependent exchange coupling constant. The potentials
$U(x),J(x,y),\,\textrm{and } V(x,y)$ vanish as $x\rightarrow \pm
\infty$. The potential $V(x,y)$ is very sharp in comparison with
the variation of $U(x)$ in the $x$-direction due to the narrowness
of the windows connecting the QPCs and the quantum-dot region.
$J(x,y)$ has an $x$-dependence similar to that of $U(x)$, since
the spatial characteristics of the local magnetic moment formed in
the conducting channel are determined by the shape of this
channel.

We write the Schr{\"o}dinger equation in the form \bq
\hat{H}_{0}\hat{\psi}(x,y)=E\hat{\psi}(x,y), \label{eq:F2} \eq
where the symbol "hat" in this and subsequent equations is used
for operators and wave functions in the four-dimensional spin
space of the two spins. The basis vectors in this space (uncoupled
representation) are given by \cite{KunzeBagwell} \bq
\hat{\chi}_{1}=\left|\uparrow_{e}\right\rangle
\left|\uparrow_{S}\right\rangle , \:
\hat{\chi}_{2}=\left|\downarrow_{e}\right\rangle
\left|\downarrow_{S}\right\rangle , 
\hat{\chi}_{3}=\left|\uparrow_{e}\right\rangle
\left|\downarrow_{S}\right\rangle, \:\textrm{and }
\hat{\chi}_{4}=\left|\downarrow_{e}\right\rangle
\left|\uparrow_{S}\right\rangle ,\label{eq:F3} \eq where
$\left|\uparrow_{e}\right\rangle$
($\left|\downarrow_{e}\right\rangle$) and
$\left|\uparrow_{S}\right\rangle$
($\left|\downarrow_{S}\right\rangle$) are spin-up (spin-down)
states of the electron spin, $\vec{\sigma}$, and the local moment
spin, $\vec{S}$, respectively. The canonical transformation to the
coupled representation is discussed in Appendix A.

The solution of the Schr\"{o}dinger equation, Eq. (\ref{eq:F2}),
can be expanded in terms of the spin functions, Eq. (\ref{eq:F3}),
as
\begin{equation}
\hat{\psi}(x,y)=\sum_{\alpha=1}^{4}\hat{\chi}_{\alpha}\psi_{\alpha}(x,y).
\end{equation}

Following the procedure of Ref. \cite{Gurvitz1} we expand the full
wave functions in terms of different propagating modes \bq
\hat{\psi}(x,y)=\sum_{n}\hat{\varphi}_{n}(x)\Phi_{n}(y) \eq  with
the transverse structure of $n^{th}$ mode given by the solutions
of the equation \bq
\left[K_{y}+W(y)\right]\Phi_{n}(y)=E_{n}\Phi_{n}(y). \eq
Correspondingly, the wave functions $\hat{\varphi}_{n}(x)$ obey
the coupled equations \bq \left[ E-E_{n}-K_{x}-U_{n}(x)\right]
\hat{\varphi}_{n}(x) =\sum_{m\neq
n}\left(V_{nm}(x)-J_{nm}(x)\hat{\vec{\sigma}}\cdot\hat{\vec{S}}\right)\hat{\varphi}_{m}(x)\label{eq:F7}
\eq where

 \bq V_{nm}(x)=\int dy \Phi_{n}^{*}(y)V(x,y)\Phi_{m}(y),
\eq \bq J_{nm}(x)=\int dy \Phi_{n}^{*}(y)J(x,y)\Phi_{m}(y),\eq and
$U_{n}(x)=U(x)+V_{nn}(x)$.

In the following analysis we make a number of simplifications in
Eq. (\ref{eq:F7}). First, we note that if the wires are well
separated, the wave functions $\Phi_{n}(y)$ are strongly localized
in one of the two wires, allowing us to distinguish the modes
propagating in each of the wires. We assume that the shape of the
confining potential $W(y)$ is such that  one of the wires is close
to pinch off (the \emph{swept} wire), i.e. it has only one
propagating mode (described by the wave function
$\hat{\varphi}_{0}(x)$) with the transverse confinement (subband
bottom) energy, $E_{0}$, less than the Fermi energy, whereas the
other wire (the \emph{fixed} wire) has several propagating modes
(Figure 3). The localized magnetic moment is supposed to form in
the only subband of the swept wire, hence the exchange coupling
can be approximated as $J_{nm}(x)=\delta_{n,0}\delta_{m,0}J(x)$.
Thus the system of equations is reduced to

\bn \left[
E-E_{0}-K_{x}-U_{0}(x)+J(x)\hat{\vec{\sigma}}\cdot\hat{\vec{S}}\right]
\hat{\varphi}_{0}(x) = \sum_{n\geq
1}V_{0n}(x)\hat{\varphi}_{n}(x)\en  and \bq
\left[E-E_{n}-K_{x}-U_{n}(x)\right] \hat{\varphi}_{n}(x)
=\sum_{m}V_{nm}(x)\hat{\varphi}_{m}(x) \text{   for  } n\geq 1.
\label{eq:F10} \eq

Furthermore, relying on the large energy separation between the
subbands, in comparison with the magnitudes of $V_{nm}(x)$ and
$J(x)$, we will neglect interaction between different subbands of
the fixed wire, effectively restricting our analysis to a
two-subband model, i.e. studying the only subband of the swept
wire and $n^{th}$ subband of the fixed wire. The coupled equations
for this pair of subbands are \bn \left[
E-E_{0}-K_{x}-U_{0}(x)+J(x)\hat{\vec{\sigma}}\cdot\hat{\vec{S}}\right]
\hat{\varphi}_{0}(x) = V_{n}(x)\hat{\varphi}_{n}(x), \en \bn
\left[E-E_{n}-K_{x}-U_{n}(x)\right] \hat{\varphi}_{n}(x)
=V_{n}(x)\hat{\varphi}_{0}(x),  \en \label{eq:F11} where we have
introduced $V_{n}(x)=V_{0n}(x)=V_{n0}(x)$.

Eqs. (12,13) can be decoupled using Green's functions: \bq
\hat{G}_{0}(\epsilon)=\left[\epsilon-K_{x}-U_{0}(x)+J(x)\hat{\vec{\sigma}}\cdot\hat{\vec{S}}\right]^{-1}
\label{eq:F12a} \eq and \bq
\hat{G}_{n}(\epsilon)=\left[\epsilon-K_{x}-U_{n}(x)\right]^{-1}.\label{eq:F12b}
\eq With these Green's functions Eqs. (12,13) can be formally
integrated as
 \bq
\hat{\varphi}_{0}(x)=\hat{G}_{0}(E-E_{0})V(x)\hat{\varphi}_{n}(x)
\nonumber \eq and \bq
\hat{\varphi}_{n}(x)=\hat{G}_{n}(E-E_{n})V(x)\hat{\varphi}_{0}(x).
\eq Accordingly, we obtain \bq \left[
E-E_{0}-K_{x}-U_{0}(x)+J(x)\hat{\vec{\sigma}}\cdot\hat{\vec{S}}\right]
\hat{\varphi}_{0}(x)
=V(x)\hat{G}_{n}(E-E_{n})V(x)\hat{\varphi}_{0}(x), \nonumber \eq
and \bq
 \left[E-E_{n}-K_{x}-U_{n}(x)\right] \hat{\varphi}_{n}(x)
=V(x)\hat{G}_{0}(E-E_{0})V(x)\hat{\varphi}_{n}(x).\label{eq:F13}
\eq

The Green's function $\hat{G}_{n}(\epsilon)$ is a scalar Green's
function, i.e. it is a unit matrix in the uncoupled spin space,
whereas $\hat{G}_{0}(\epsilon)$ has a more complicated structure.
Nevertheless, it can be expressed in terms of two scalar Green's
functions (see the derivation in Appendix B) as \bq
\hat{G_0}(\epsilon)=\frac{1}{4}\left[3g^{t}(\epsilon)+g^{s}(\epsilon)\right]\hat{I}+
\frac{1}{4}\left[g^{t}(\epsilon)-g^{s}(\epsilon)\right]\hat{\vec{\sigma}}\cdot\hat{\vec{S}},\eq
where \bq g^{t}(\epsilon)=\left[ \epsilon
-K_{x}-U(x)+J(x)\right]^{-1} \label{eq:GG2a}\eq and \bq
g^{s}(\epsilon)=\left[ \epsilon
-K_{x}-U(x)-3J(x)\right]^{-1}.\label{eq:GG2b}
 \eq
Now we are able to redefine the scalar potentials, as \bq
\tilde{U}_{0}(x,E)=U_{0}(x)+V(x)\hat{G}_{n}(E-E_{n})V(x)
\eq\label{eq:F15} and \bq \tilde{U}_{n}(x,E)=U_{n}(x) +
v_n(x,E)=U_{n}(x) +
V_{n}(x)\frac{1}{4}\left[3g^{t}(E-E_{0})+g^{s}(E-E_{0})\right]V_{n}(x),\label{eq:F16}
\eq and introduce the tunneling-induced exchange coupling of
electrons in the fixed wire to the local magnetic moment, \bq
j_{n}(x,E)=-V_{n}(x)\frac{1}{4}\left[g^{t}(E-E_{0})-g^{s}(E-E_{0})\right]V_{n}(x).\label{eq:F17}\eq
As a result, we obtain the following equations for the description
of electron dynamics in the swept and fixed wires in the form \bq
\left[
E-E_{0}-K_{x}-\tilde{U}_{0}(x)+J(x)\hat{\vec{\sigma}}\cdot\hat{\vec{S}}\right]
\hat{\varphi}_{0}(x)=0 , \label{eq:F18a}\eq and \bq
\left[E-E_{n}-K_{x}-\tilde{U}_{n}(x)+j(x,E)\hat{\vec{\sigma}}\cdot\hat{\vec{S}}\right]
\hat{\varphi}_{n}(x) =0.\label{eq:F18b} \eq

Although the form of these two equations is very similar, and they
can be both treated in the same manner (as is discussed in the
next Section), the results they yield will differ, depending on
the specific shapes of the potentials and exchange couplings. In
particular, while the shape of the coupling $J(x)$ in Eq.
(\ref{eq:F18a}) is smooth, similar to that of the potential
$U(x)$, the exchange constant $j(x)$ of Eq. (\ref{eq:F18b}) is
proportional to the potential $V(x)$, and therefore is sharper
than the bottleneck potential $U(x)$.

\section{Calculations of the transmission coefficient and conductance for the swept and fixed wires} \label{sec:08}

In our previous paper \cite{IEEE}, we determined the transmission
coefficient and the conductance of a single QPC, expanding
functions $\tilde{U}_{0}(x)$ and $J(x)$ involved in Eq.(26) into
series near their maxima (i.e. representing them as inverted
parabolas) as

 \bq
 \tilde{U}_{0}(x)=\tilde{U}_{0}(0)+\frac{x^2}{2}\frac{\partial^{2}\tilde{U}_{0}(x)}{\partial
 x^{2}}|_{x=0}=\tilde{U}_{max}-\frac{m\omega_{U}^{2}x^{2}}{2} \label{eq:SW1a}\eq and \bq
  J(x)=J(0)+\frac{x^2}{2}\frac{\partial^{2}J(x)}{\partial
 x^{2}}|_{x=0}=J_{max}-\frac{m\omega_{J}^{2}x^{2}}{2}.
 \label{eq:SW1b} \eq

The transmission coefficient for the inverse parabolic barrier
$u(x)=-m\omega^{2}x^{2}/2$ is given by \cite{Levinson} \bq
t(\eta)=\left[1+e^{-2\pi\eta}\right]^{-1/2}, \eq where
$\eta=\epsilon/\hbar \omega$, and the energy, $\epsilon$, is
measured from the top of the barrier. Thus, the transmission
coefficients of the swept wire can be written as \bq
T_{0t}=t\left(\frac{\epsilon-\tilde{U}_{max}+J_{max}}{\hbar
\omega_{-}}\right)\eq and \bq
T_{0s}=t\left(\frac{\epsilon-\tilde{U}_{max}-3J_{max}}{\hbar
\omega_{+}}\right) ,\eq where
$\omega_{-}=\sqrt{\omega_{U}^{2}-\omega_{J}^{2}}$,
 $\omega_{+}=\sqrt{\omega_{U}^{2}+3\omega_{J}^{2}}$.
Assuming the equivalence of all initial spin orientations, we
obtain the conductance of the swept wire as \bn
G_{SW}=\frac{2e^2}{h}\left[\frac{3}{4}\left|T_{0t}\right|^{2}
+\frac{1}{4}\left|T_{0s}\right|^{2}\right]= \nonumber \\
\frac{2e^{2}}{h}\left[\frac{3}{4}\left|t\left(\frac{\epsilon-\tilde{U}_{max}+J_{max}}{\hbar
\omega_{-}}\right)\right|^2
+\frac{1}{4}\left|t\left(\frac{\epsilon-\tilde{U}_{max}-3J_{max}}{\hbar
\omega_{+}}\right)\right|^{2}\right] . \en

 The most important feature of the
transmission coefficients is that the transmission probability,
$\left|t(\eta)\right|^{2}$, is very close to a step function. This
step-like structure causes the conductance to reproduce the
step-like behavior of the 0.7-anomaly. In the case of
ferromagnetic coupling between the electrons and local magnetic
moment, $J_{max}>0$, our model gives an additional conductance
step at $0.75\times2e^{2}/h$, as  \bq
G_{SW}=\frac{2e^{2}}{h}\left\{\begin{array}{ll}
  0,& \textrm{if } \epsilon<\tilde{U}_{max}-J_{max},  \\
  0.75, & \textrm{if } \tilde{U}_{max}-J_{max}<\epsilon<\tilde{U}_{max}+3J_{max}, \\
  1 ,& \textrm{if } \epsilon >\tilde{U}_{max}+3J_{max}. \\
\end{array}\right.
\eq It is interesting to point out that for antiferromagnetic
coupling, $J_{max}<0$, we obtain a conductance step at $0.25\times
2e^2/h$, which has been observed in experiments
 \cite{future} and density-functional simulations \cite{Berggren},
as
 \bq
G_{SW}=\frac{2e^{2}}{h}\left\{\begin{array}{ll}
  0,& \textrm{if } \epsilon<\tilde{U}_{max}-3\left|J_{max}\right|,  \\
  0.25, & \textrm{if } U_{max}-3\left|J_{max}\right|<\epsilon<\tilde{U}_{max}+\left|J_{max}\right|, \\
  1 ,& \textrm{if } \epsilon >\tilde{U}_{max}+\left|J_{max}\right|. \\
\end{array}\right.
\eq

The idea that the 0.7-anomaly is caused by singlet-triplet
splitting of the first plateau, into the triplet part contributing
3/4(=0.75) and the singlet part contributing 1/4(=0.25,) was
suggested in Refs. \cite{Rejec} and \cite{Flambaum}. However,
these theories failed to reproduce the correct behavior of
0.7-anomaly with variations of temperature, concentration and
source-drain bias. Correspondingly, the model used in the present
paper is also too primitive but it can be improved, in particular,
by using the results of density functional modelling
\cite{Berggren,Hirose1} to specify the shape and strength of the
exchange coupling $J(x,y)$ by comparing phenomenological
parameters to experimental data. It should be also noted that in
experiments the actual position of the "0.7-plateau" varies
between 0.5 and 0.8 for samples having different electron
concentrations, gate voltages, and source-drain biases (see
\cite{Science} and references therein) and, accordingly, the
theoretical explanations providing the "0.75" result cannot be
ruled out especially in a view of experimental observation of the
0.25-plateau \cite{future}.

The method of calculation of the {\it fixed wire} conductance is
very similar to that of the swept wire. However, the exchange and
scattering potentials, involved in Eq. (26) for the swept wire and
in Eq. (27) for the fixed wire, are different, leading to the
differences in the behavior of the conductance. One of the main
differences is that the tunneling channel, whose width is
characterized by the width of potential $V_{n}(x)$, is narrow in
comparison to the extent of the bottleneck potential of a quantum
wire, described by $U_{n}(x)$, and the corrections associated with
this tunneling appear as a peak or a dip on top of the potential.

To evaluate Eq. (27), we rewrite it in the coupled representation
(see appendix A, with the prime to be omitted below) as \bq \left(
E-E_n
-K_{x}-U_{n}(x)-v_{n}(x)+j_{n}(x,E)\right)\varphi_{n\alpha}(x)=0\eq
for $\alpha=1,2,3$ and \bq \left( E-E_n
-K_{x}-U_{n}(x)-v_{n}(x)-3j_{n}(x,E)\right)\varphi_{n4}(x)=0\eq
for $\alpha=4$. The exchange-independent solutions can be found
from the equation \bq \left( E-E_n
-K_{x}-U_{n}(x)-v_{n}(x)\right)\chi_{nk}^{\pm}(x)=0, \eq where
$k=\frac{1}{\hbar}\sqrt{2m(E-E_n)}$, and we denote the
transmission coefficient associated with these solutions as
$t_n(E-E_n)$.

We can express the exchange term, $j_n(x,E)$ in terms of the
transmission coefficients of the swept wire. In this, we employ
the approximation of inverse parabolicity of the barrier in the
swept wire to the Green's functions involved in the definition of
the exchange term, \bq j_{n}(x)=-V_{n}(x)\frac{1}{4}\int dx'
\left[g^{t}(x,x',E-E_{0})-g^{s}(x,x',E-E_{0})\right]V_{n}(x').\eq
Using the properties of the Green's functions of the inverse
parabolic barrier (see Appendix C), we find that the energy
dependence of the exchange term is determined by the difference of
the transmission coefficients as \bq j_{n}(x,E)\sim \left[
T_{0t}(E-E_{0})-T_{0s}(E-E_{0})\right]j_{n}(x). \eq

The contributions of the exchange interaction have different signs
for the singlet and triplet states and appear as a peak and a dip,
respectively, on top of the bottleneck potential
$\tilde{U}(x)=U(x)+v(x)$ (see Figure 4). The dip leads to the
occurrence of localized states inside the potential of the fixed
wire modifying its conductive properties. We consider two possible
situations.

1. Ferromagnetic coupling ($j_n(x,E)>0$).

In this case the triplet states experience a dip in the potential,
and the energy of the quasibound state (Figure 4) can be found
from the equation \bq \left[ K_{x}-j_{n}(x,E)
\right]\phi_{nt}(x)=\lambda_{nt}\phi_{nt}(x), \eq where the energy
is counted from the top of the bottleneck potential,
$\tilde{U}_{n,max}$, and $\lambda_{nt}$ is negative. The
transmission coefficient of a barrier with the quasibound state
was calculated in Ref. \cite{Gurvitz1}, and is given by \bq
T_{nt}(E-E_n)=t_{n}(E-E_n)+\frac{m}{\imath k \hbar^{2}}
\frac{\left\langle \phi_{nt}\left| j_{n}(x,E)\right|
\chi_{nk}^{+}\right\rangle\left\langle \phi_{nt}\left|
 j_{n}(x,E)\right| \chi_{nk}^{-}\right\rangle
}{E-E_n-\tilde{U}_{n,max}-\bar{\lambda}_{nt}+\imath \Gamma_{nt}},
\eq where $\bar{\lambda}_{nt}=\lambda_{nt}+\delta\lambda_{nt}$
(with $\delta\lambda_{nt}$ accounting for the energy shift due to
the possibility of tunneling in and out of the quasibound state)
and the width of the tunneling resonance, $\Gamma_{nt}$, has the
form \cite{Gurvitz1} \bq
\Gamma_{nt}=\frac{m}{2k\hbar^{2}}\left(\left|\left\langle
\phi_{nt}\left| j_{n}(x,E)\right|
\chi_{nk}^{+}\right\rangle\right|^{2}+\left|\left\langle
\phi_{nt}\left| j_{n}(x,E)\right|
\chi_{nk}^{-}\right\rangle\right|^{2}\right). \eq Substituting the
expressions for the exchange term, Eq. (40), into Eq. (42), we
obtain \bq T_{nt}(E-E_{n})=t_{n}(E-E_{n})+ \frac{K_{n}\left[
T_{0t}(E-E_{0})-T_{0s}(E-E_{0})\right]^{2}
}{E-E_{n}-\tilde{U}_{n,max}-\bar{\lambda}_{nt}+\imath
\Gamma_{nt}(E-E_{0})}, \label{eq:61}\eq where $K_n$ is a scalar
coefficient. The bottleneck potential of the fixed wire can also
be assumed to be inverse parabolic,
$\tilde{U}_{n}(x)\approx\tilde{U}_{n,max}-m\Omega_{n}^{2}x^{2}/2$,
and the background transmission coefficient has the form \bq
t_{n}(E-E_{n})=t\left[\frac{E-E_{n}-\tilde{U}_{n,max}}{\hbar\Omega_{n}}\right].
\label{eq:62} \eq The absolute value of the transmission
coefficient Eq. (\ref{eq:61}) should not exceed unity. One can see
that this condition is obeyed because the two terms in this
expression are non-zero for different energies.

For the singlet state there is no dip in the potential, but the
barrier is a little bit higher than $\tilde{U}_{n,max}$, which can
be taken into account by introducing parameter $\delta
j_{n}(E-E_{0})$ proportional to $\left[
T_{0t}(E-E_{0})-T_{0s}(E-E_{0})\right]\delta\tilde{j}_{n}$, so
that the transmission coefficient for the singlet state can be
written as \bq
T_{ns}(E-E_{n})=t\left[\frac{E-E_{n}-\tilde{U}_{n,max}-\delta
j_{n}(E-E_{0})}{\hbar\Omega_{n}}\right]. \eq Finally, the width of
the tunneling resonance takes form \bq
\Gamma_{nt}(E-E_{0})=\Gamma_{n,0} \left[
T_{0t}(E-E_{0})-T_{0s}(E-E_{0})\right]^{2} ,\eq where
$\Gamma_{n,0}$ is a constant.

2. Antiferromagnetic coupling ($j_n(x)<0$).

In this case the singlet state experiences scattering through
quazibonding state, whose bare energy and zero-order wave function
can be determined by the equation \bq \left[ K_{x}+3j_{n}(x,E)
\right]\phi_{ns}'(x)=\lambda_{ns}'\phi_{ns}'(x). \eq Employing the
same procedure as in the previous case, we obtain the transmission
coefficients as \bq T_{ns}'(E-E_{n})=t_{n}(E-E_{n})+
\frac{K_{n}'\left[ T_{0t}(E-E_{0})-T_{0s}(E-E_{0})\right]^{2}
}{E-E_{n}-\tilde{U}_{n,max}-\bar{\lambda}_{ns}'+\imath
\Gamma_{ns}'(E-E_{0})}, \eq \bq
T_{nt}'(E-E_{n})=t\left[\frac{E-E_{n}-\tilde{U}_{n,max}-\delta
j_{n}'(E-E_{0})}{\hbar\Omega_{n}}\right]. \eq In these expressions
 \bq
\Gamma_{ns}'(E-E_{0})=\Gamma_{n,0}' \left[
T_{0t}(E-E_{0})-T_{0s}(E-E_{0})\right]^{2}. \eq We can establish
approximate relations between the coefficients in the
ferromagnetic and antiferromagnetic cases: $K_{n}'\approx 9K_{n}$,
$\Gamma_{n}'\approx 9 \Gamma_{n}$, and $\delta j_{n}'\approx 3
\delta j_{n}$.

With these transmission coefficients we can obtain an expression
for the conductance as  \bq
G_{FW}=\frac{2e^{2}}{h}\sum_{n}\left[\frac{3}{4}\left|T_{nt}(E-E_{n})\right|^2  
+\frac{1}{4}\left|T_{ns}(E-E_{n})\right|^{2}\right] . \eq The
conductances of the swept and fixed wires is shown in Figure 5 as
functions of the gate voltage of the swept wire (which determines
the energy separation of the local state, $E_0$, and the Fermi
energy) for the ferromagnetic case and for the following set of
parameters: $E_F-\tilde{U}_{max} = 0.6 meV, \tilde{U}_{max}-E_n =
0.3 meV, J_{max} = 0.3 meV, \omega_- = 0.3 meV, \omega_+ = 1.5
meV, \omega_U = 1 meV, \Omega_n = 1 meV, K_n = 0.0285 meV,
\Gamma_n = 0.1 meV,$ and $\delta j_n = 0.1 meV$. The confinement
potential in the fixed wire is assumed to be parabolic with the
level separation $E_n-E_{n-1} = 0.3 meV$.

One can see from this figure that the conductance peak in the
fixed wire appears exactly at the same gate voltages as the
0.75-plateau in the conductance of the swept wire indicating their
common nature as the local moment formation as the swept wire
pinches off.

\section{Conclusions}

In this report, we have presented a comprehensive theory for the
electron dynamics in a system of coupled quantum wires, under
conditions where a local magnetic moment is formed in one of them.
Rather than assume that this local moment is related to the
formation of an associated localized state in the swept wire, we
have calculated the single-electron transmission properties of the
fixed wire in a potential that is modified by the presence of an
extra scattering term, arising from the presence of the local
moment in the swept wire. To determine the transmission
coefficients in this system, we derived equations describing the
dynamics of electrons in the swept and fixed wires of the
coupled-wire geometry. Our analysis clearly shows that the
observation of a resonant peak in the conductance of the fixed
wire is correlated to the appearance of additional structure (near
$0.75\cdot$ or $0.25\cdot 2e^2/h$) in the conductance of the swept
wire, in agreement with the experimental results of Ref. [6].

\vspace*{1cm}
 {\large {\bf Acknowledgement}}

The authors would like to thank S. A. Gurvitz, A. Yu. Smirnov and
I. M. Djuric for the useful discussion and critical comments.

\appendix

\section{Electron scattering by a localized spin}
\label{Appx:Q}

In this appendix we discuss the canonical transformation from the
uncoupled representation to the coupled representation.

The basis vectors in  the spin space of electron spin and the
magnetic moment in the uncoupled representation are given by Eqs.
(\ref{eq:F3}). The form of the exchange operator
$\vec{\sigma}\cdot\vec{S}$ in this basis is
\begin{equation}
\hat{Q}=\vec{\sigma}\cdot\vec{S}=\left(%
\begin{array}{cccc}
  1 & 0 & 0 & 0 \\
  0 & 1 & 0 & 0 \\
  0 & 0 & -1 & 2 \\
  0 & 0 & 2 & -1
\end{array}%
\right).\label{eq:Q1}
\end{equation}
This operator can be diagonalized by a canonical transformation
\begin{equation}
\hat{Q}'=\hat{X}^{+}\hat{Q}\hat{X}=\left(%
\begin{array}{cccc}
  1 & 0 & 0 & 0 \\
  0 & 1 & 0 & 0 \\
  0 & 0 & 1 & 0 \\
  0 & 0 & 0 & -3 \\
\end{array}%
\right)\label{eq:Q2}
\end{equation}
where the transformation operator is given by
\begin{equation}
\hat{X}=\left(%
\begin{array}{cccc}
  1 & 0 & 0 & 0 \\
  0 & 1 & 0 & 0 \\
  0 & 0 & \frac{1}{\sqrt{2}} & -\frac{1}{\sqrt{2}} \\
  0 & 0 & \frac{1}{\sqrt{2}} & \frac{1}{\sqrt{2}} \\
\end{array}%
\right) \label{eq:Q3}
\end{equation}
The wave function is transformed in a similar way:
\begin{equation}
\hat{\varphi}'(x)=\hat{X}^{+}\hat{\varphi}(x). \label{eq:Q5}
\end{equation}

The equation describing scattering of an electron on LMM is
\begin{equation}
\left( \epsilon -K_{x}-U(x)+J(x)\vec{\sigma}\cdot \vec{S}
\right)\varphi(x)=0, \label{eq:Q5}
\end{equation}
where $\varphi(x)$ is a four-component wave function in spin
space:
\begin{equation}
\hat{\varphi}(x)=\sum_{\alpha=1}^{4}\chi_{\alpha}\varphi_{\alpha}(x).
\end{equation}
This equation can be formally solved with help of the canonical
transformation, Eq. (\ref{eq:Q3}) ($\hat{I}=\hat{X}^{+}\hat{X}$):
\bn \hat{X}^{+}\left( \epsilon -K_{x}-U(x)+J(x)\vec{\sigma}\cdot
\vec{S} \right)\hat{X}\hat{X}^{+}\varphi(x) \nonumber \\=\left(
\epsilon -K_{x}-U(x)+J(x)\hat{X}^{+}\vec{\sigma}\cdot \vec{S}
\hat{X}\right)\hat{\varphi}'(x)\nonumber \\=\left( \epsilon
-K_{x}-U(x)+J(x)\hat{Q}'\right)\hat{\varphi}'(x)=0. \label{eq:Q7}
\en Equation (\ref{eq:Q7}) is diagonal in spin space and can be
written as four equations for wave function components:\bq \left(
\epsilon -K_{x}-U(x)+J(x)\right)\varphi'_{\alpha}(x)=0,\,
\alpha=1,2,3 \label{eq:Q8a}\eq \bq \left( \epsilon
-K_{x}-U(x)-3J(x)\right)\varphi'_{4}(x)=0.\label{eq:Q8b}
 \eq\label{eq:Q8}

\section{Green's function for an electron scattered by a localized
spin}\label{Appx:G} In this Appendix we derive the Green's
function of Eq. (20), starting from its definition, Eq.
(\ref{eq:F12a}),\bq
\hat{G}_{0}(\epsilon)=\left[\epsilon-K_{x}-U_{0}(x)+J(x)\hat{\vec{\sigma}}\cdot\hat{\vec{S}}\right]^{-1}\nonumber
\eq According to this, the Green's function satisfies the equation
\bq \left( \epsilon -K_{x}-U(x)+J(x)\hat{\vec{\sigma}}\cdot
\hat{\vec{S}} \right)\hat{G}(x,x',\epsilon)=\hat{I}\delta(x-x'),
\eq where $\hat{I}$ is the unit matrix in spin space, together
with the boundary conditions, which depend on the particular kind
of the Green's function that we are looking for.

Using the canonical transformation of Appendix \ref{Appx:Q}, we
can calculate the Green's function
$\hat{G}'(x,x',\epsilon)=\hat{X}^{+}\hat{G}(x,x',\epsilon)\hat{X}$,
which is diagonal in spin space,
$G_{\alpha\beta}'(x,x',\epsilon)=\delta_{\alpha\beta}G_{\alpha}'(x,x',\epsilon)$,
and whose components,
$G_{\alpha}'(x,x',\epsilon)=g^{t}(x,x',\epsilon)\textrm{ for }
\alpha=1,2,3$, and $G_{4}'(x,x',\epsilon)=g^{s}(x,x',\epsilon)$,
should satisfy the equations

 \bq \left( \epsilon
-K_{x}-U(x)+J(x)\right)g^{t}(x,x',\epsilon)=\delta(x-x'),
\label{eq:G2a}\eq \bq \left( \epsilon
-K_{x}-U(x)-3J(x)\right)g^{s}(x,x',\epsilon)=\delta(x-x').\label{eq:G2b}
 \eq\label{eq:G2}

The outgoing-wave Green's function,
$G_{\alpha\beta}(x,x',\epsilon)\sim \delta_{\alpha,\beta}e^{\pm
ikx}$ for $x\longrightarrow \pm \infty$, is of most interest to
us. It is shown in Ref. \cite{Gurvitz1} that in terms of the
scattering solutions the components of the desired Green's
function are given by \bq
g^{t,s}(x,x',\epsilon)=\frac{m}{ikT_{t,s}}\left\{\begin{array}{cc}
  \phi_{k}^{t,s-}(x')\phi_{k}^{t,s+}(x) & \textrm{if } x>x', \\
  \phi_{k}^{t,s+}(x')\phi_{k}^{t,s-}(x) & \textrm{if } x<x', \\
\end{array}\right.
\eq where $k=\sqrt{2m\epsilon}/\hbar$ and $\phi_{k}^{t,s-/+}$ are
triplet/singlet scattering solutions originated from +/- $\infty$,
respectively.

Now the Green's function $\hat{G}'(x,x',\epsilon)$ takes form \bq
\hat{G}'(x,x',\epsilon)=\left(%
\begin{array}{cccc}
  g^{t}(x,x',\epsilon) & 0 & 0 & 0 \\
  0 & g^{t}(x,x',\epsilon) & 0 & 0 \\
  0 & 0 & g^{t}(x,x',\epsilon) & 0 \\
  0 & 0 & 0 & g^{s}(x,x',\epsilon) \\
\end{array}%
\right) \eq and applying the canonical transformation backwards we
obtain \bn
\hat{G}(x,x',\epsilon)=\hat{X}\hat{G}'(x,x',\epsilon)\hat{X}^{+}= \nonumber \\ \left(%
\begin{array}{cccc}
  g^{t}(x,x',\epsilon) & 0 & 0 & 0 \\
  0 & g^{t}(x,x',\epsilon) & 0 & 0 \\
  0 & 0 & \frac{1}{2}\left[g^{t}(x,x',\epsilon)+g^{s}(x,x',\epsilon)\right] & \frac{1}{2}\left[g^{t}(x,x',\epsilon)-g^{s}(x,x',\epsilon)\right] \\
  0 & 0 & \frac{1}{2}\left[g^{t}(x,x',\epsilon)-g^{s}(x,x',\epsilon)\right] & \frac{1}{2}\left[g^{t}(x,x',\epsilon)+g^{s}(x,x',\epsilon)\right] \\
\end{array}%
\right) \label{eq:G5}\en

Finally, Eq. (\ref{eq:G5}) can be split into the scalar part,
proportional to a unit matrix in the spin space, and the part
proportional to the exchange operator,
$\hat{Q}=\hat{\vec{\sigma}}\cdot \hat{\vec{S}}$:

\bq
\hat{G}(x,x',\epsilon)=\frac{1}{4}\left[3g^{t}(x,x',\epsilon)+g^{s}(x,x',\epsilon)\right]\hat{I}+
\frac{1}{4}\left[g^{t}(x,x',\epsilon)-g^{s}(x,x',\epsilon)\right]\hat{\vec{\sigma}}\cdot\hat{\vec{S}}.\eq

\section{Green's functions and transmission coefficients of an inverse parabolic barrier}
In this section we briefly summarize some known facts about an
inverse parabolic barrier \cite{Levinson}. If the barrier
potential is given by $u(x)=-m\omega^{2}x^{2}/2$, the scattering
solutions of the equation \bq
\left[\epsilon-K_{x}-u(x)\right]\Psi^{\pm}(x)=0 \eq are given by
\bq \Psi^{\pm}(x)=\mathrm{E}(\eta,\pm\xi),\label{eq:A2}\eq where
$\mathrm{E}(\eta,\xi)$ is a Weber function, i.e. a solution of the
equation for parabolic cylinder functions,
$y''(\xi)+(\frac{1}{4}\xi^{2}-\eta)y(\xi)=0$, \cite{Abramowitz};
$\xi=qx$, and $q=\sqrt{2m\omega/\hbar}$, whereas
$\eta=-\epsilon/(\hbar\omega)$.

The one-dimensional Green's function for such a barrier is given
by \bq
G(x,x',\epsilon)=\frac{mt(\eta)}{\hbar^{2}q}\left\{\begin{array}{c}
  \mathrm{E}(\eta,\xi)\mathrm{E}(\eta,-\xi'), x>x' \\
  \mathrm{E}(\eta,-\xi)\mathrm{E}(\eta,\xi'), x<x'  \\
\end{array} \right., \label{eq:A3} \eq
where the transmission coefficient has the form
 \bq
t(\eta)=\left[1+e^{-2\pi\eta}\right]^{-1/2}. \label{eq:A4} \eq

\end{document}